\documentclass[12pt]{article}

\usepackage{graphicx}
\usepackage{dcolumn}
\usepackage{bm}
\usepackage{amsmath}
\usepackage{epstopdf}
\usepackage{amsfonts}
\usepackage{amssymb}
\usepackage{epsfig}
\usepackage{tabularx}
\usepackage{color}

\usepackage[colorlinks=true,citecolor=blue,urlcolor=magenta,breaklinks]{hyperref}

\newcommand{\be}{\begin{equation}}
\newcommand{\en}{\end{equation}}
\newcommand{\bea}{\begin{eqnarray}}
\newcommand{\ena}{\end{eqnarray}}

\begin{document}

\begin{titlepage}



\centerline{\large \bf {Building (1+1) holographic superconductors in}}

\centerline{\large \bf {the presence of non-linear Electrodynamics}}

\vskip 1.5cm

\centerline{Grigoris Panotopoulos}

\vskip 1cm

\centerline{Centro de Astrof{\'i}sica e Gravita\c c\~ao-CENTRA, Departamento de F{\'i}sica,} 

\centerline{Instituto Superior T{\'e}cnico-IST, Universidade de Lisboa-UL,}

\centerline{Avenida Rovisco Pais 1, 1049-001, Lisboa, Portugal}

\vskip 1cm

\centerline{email:
\href{mailto:grigorios.panotopoulos@tecnico.ulisboa.pt}
{\nolinkurl{grigorios.panotopoulos@tecnico.ulisboa.pt}}
}

\vskip 1.5cm


\begin{abstract}
In the framework of the gauge/gravity duality, and in particular of the $AdS_3/CFT_2$ correspondence, we study one-dimensional superconductors analyzing the dual (1+2)-dimensional gravity in the presence of the Einstein-power-Maxwell non-linear Electrodynamics. In the probe limit we compute the critical temperature of the transition as a function of the mass of the scalar field. The computation is performed analytically employing the Rayleigh-Ritz variational principle. The comparison with a power Maxwell field in higher dimensions as well as with the (1+3)-dimensional Einstein-Maxwell theory for two-dimensional superconductors is made as well.
\end{abstract}


\end{titlepage}

\section{Introduction}

The Superconductor-Normal transition is one of the most exciting research areas in condensed matter physics. Superconductivity was discovered in the Leiden laboratories by H.~K.~Onnes in 1911 \cite{onnes}, when he 
observed that the electrical resistivity of mercury dropped at an unmeasurably low value at a transition 
temperature of $T_c \simeq 4~K$. The microscopic theory that explains the underlying mechanism was finally 
formulated in 1957 by Bardeen, Cooper and Schrieffer (BCS) \cite{BCS1,BCS2}. In 1986-1987, however, new 
superconducting materials were designed in which the transition temperature was found to be significantly 
larger than what was expected from the BCS theory \cite{highTc,wu}. In particular, in conventional superconductors 
the highest known transition temperature is $T_c=23.2~K$ in $Nb_3Ge$, while in the high-$T_c$ oxides the observed 
transition temperature was in the $T_c=30~K$ range in 1986 in the Ba-La-Cu-O system \cite{highTc}, and 
$T_c \simeq 90~K$ in the following year in the Y-Ba-Cu-O system \cite{wu}. The discoveries of low and high 
temperature superconductors as well as the formulation of the BCS theory have been awarded with the Nobel prize 
in Physics in 1913 \cite{NP1}, in 1987 \cite{NP2} and in 1972 \cite{NP3}, respectively.

\smallskip

One of the most remarkable consequences of Superstring Theory \cite{ST1,ST2} has been the AdS/CFT correspondence \cite{maldacena}, which was later extended to gauge/gravity duality \cite{klebanov}. The main idea is that a 
strongly coupled conformal field theory in $D$ dimensions can be understood by solving a weakly coupled 
gravitational system in $D+1$ dimensions. The aforementioned equivalence raises the hope and the expectation 
that strongly coupled condensed matter systems may be explained by black hole physics. The original conjecture 
posits an equivalence between type IIB string theory on $AdS_5 \times S_5$ and a supersymmetric $\mathcal{N}=4$ 
Yang-Mills $SU(N)$ theory in (1+3) dimensions. There are at least two arguments pointing to this kind of equivalence: 
First symmetry counting, namely a conformal theory in $(1+3)$ dimensions has $15$ degrees of freedom, while anti-de 
Sitter in $(1+4)$ dimensions has the isometries of $SO(1,4)$ with 15 generators. In addition, a stuck of $N$ parallel 
D3-branes \cite{Dbranes1,Dbranes2} can be viewed in two different ways as follows: On the one hand it naturally 
supports a four-dimensional supersymmetric gauge field theory based on $SU(N)$ gauge group with $\mathcal{N}=4$ supersymmetric generators \cite{duality}. On the other hand the stuck of D-branes generates a gravitational field 
which in the near-horizon limit becomes $AdS_5 \times S_5$ \cite{duality}. 

\smallskip

High-$T_c$ superconductivity cannot be described by the BCS theory, and it is one of the most enigmatic areas of 
condensed matter physics. The pioneer works of \cite{HS1,HS2} in 2008 marked the birth of the field of holographic superconductors, and by now it is a very active one. For reviews see e.g. \cite{review1,review2}. To build holographic superconductors the minimal ingredients are the following: a) a gravitational theory with a negative cosmological constant, b) the Maxwell potential $A_\mu$ and c) an electrically charged massive scalar field \cite{duality}. However, in principle the electromagnetic theory may be a non-linear one, such as Born-Infeld (BI) or the so called Einstein-power-Maxwell (EpM) theory. 

\smallskip

Non-linear Electrodynamics (NLE), with already a long history, has attracted a lot of attention, and it has been studied over the years in several different contexts. As a starting point one may mention that classical electrodynamics consists of a system of linear equations, at quantum level however, when radiative corrections are taken into account, the effective equations inevitably become non-linear. The first works go back to the 30's, when Euler and Heisenberg managed to obtain 
QED corrections \cite{Euler}, while Born and Infeld were able to obtain a finite self-energy of point-like charges \cite{BI}. Moreover, a generalization of Maxwell's theory leads to the so called Einstein-power-Maxwell (EpM) theory \cite{EpM1,EpM2,EpM3,EpM4,EpM5,EpM6,EpM7,EpM8,EpM9,EpM10,EpM11}, described by 
a simple Lagrangian density of the form $\mathcal{L}(F) = F^q$, with $F$ being the Maxwell invariant, and $q$ being an arbitrary rational number. This class of NLE preserves the nice properties of conformal invariance in any number of space time dimensionality $D$, provided that $q=D/4$. Furthermore, assuming appropriate non-linear electromagnetic sources, which in the weak field limit boil down to Maxwell's linear theory, one may generate new solutions (Bardeen-like solutions \cite{Bardeen}, see also \cite{borde}) to Einstein's field equations \cite{beato1,beato2,beato3,bronnikov,dymnikova,hayward,vagenas1,vagenas2}. Those new solutions on the one hand have a horizon, and on the other hand their curvature invariants, such as the Ricci scalar $R$, are regular everywhere. This
is to be contrasted to the usual Reissner-Nordstr{\"o}m solution \cite{RN}, which possesses a singularity at the origin,
$r \rightarrow 0$.

\smallskip

Holographic superconductors in the presence of EpM theory have been studied e.g. in \cite{NLE1,davood1,NLE2,NLE3} for $D \geq 4$ (see also \cite{ruth1,lefteris,ruth2} for works on holographic superconductors in the Einstein-Gauss-Bonnet gravity 
in higher dimensions, and \cite{davood2,davood3} for backreaction and entanglement entropy in three-dimensional Einstein-Maxwell theory), and in the presence of BI in \cite{NLE4,NLE5,NLE6,NLE7,NLE8,NLE9,NLE10,NLE11,NLE12}. For additional interesting articles in the context of BI and a power Maxwell field, which motivate even further the studies on non-linear Electrodynamics, and also in connection to the Ba{\~n}ados-Teitelboim-Zanelli (BTZ) black hole \cite{BTZ1,BTZ2} and its extensions in several different contexts (geodesics in massive gravity, solutions with a scalar hair, exact solutions in General Relativity versus $f(R)$ theories, dilatonic black holes etc), see
\cite{extra1,extra2,extra3,extra4,extra5,extra6,extra7, extra8, extra9, extra10}.

In the present work we propose to build one-dimensional holographic superconductors in the presence of EpM NLE, 
which to the best of our knowledge is still missing, filling thus a gap in the literature. Our work is 
organized as follows: In the next section we present the model and the field equations, while the critical temperature of the transition is discussed in section 3, where our numerical results are shown. Finally, we conclude in the last section.

\section{Model and field equations}

First we shall keep the discussion as general as possible for any power $q$ and dimensionality of space-time $D$, 
following \cite{NLE1,NLE2}, and then we shall consider a few special cases.

We consider a gravitational system described by the action
\begin{eqnarray}
S & = & S_G + S_M  \\
S_G & = & \int d^Dx \sqrt{-g} \left(\frac{R}{16 \pi G} - 2 \Lambda \right) \\
S_M & = & \int d^Dx \sqrt{-g} [\mathcal{L}(F)_{EM} + |D_\mu \psi|^2 - m^2 |\psi|^2]
\end{eqnarray}
where the gravitational part, $S_G$, includes the Einstein-Hilbert term and a negative cosmological constant, $\Lambda=-(D-1) (D-2)/(2 l^2)$, while the matter part, $S_M$, consists of a massive charged scalar field with mass $m$, and the 
(non-linear) electromagnetic theory $\mathcal{L}(F)_{EM}=-\beta F^q$. The covariant derivative $D_\mu$ and the Maxwell's invariant $F$ are given by
\begin{equation}
D_\mu = \partial_\mu -i e A_\mu
\end{equation}
\begin{equation}
F = F_{\mu \nu} F^{\mu \nu}
\end{equation}
\begin{equation}
F_{\mu \nu} = \partial_\mu A_\nu - \partial_\nu A_\mu
\end{equation}
where $A_\mu$ is the Maxwell potential, and $e$ is the electric charge of the scalar field. 

In the following we work in the probe limit neglecting the back reaction of the matter fields on the geometry. 
At least for temperatures close to the transition temperature this should be a good approximation. We thus consider 
a fixed gravitational background of the form
\begin{equation}
ds^2 = -f(r) dt^2 + f(r)^{-1} dr^2 + r^2 dx_i dx^i
\end{equation}
which is the higher-dimensional version of the BTZ black hole \cite{BTZ1,BTZ2}, and
where the lapse function $f(r)$ is given by (setting $l=1$)
\begin{equation}
f(r) = r^2 \left[ 1 - \left( \frac{r_H}{r} \right)^{D-1}  \right]
\end{equation}
with $r_H$ being the event horizon. 

\smallskip

Varying the action with respect to the scalar field and the Maxwell potential we obtain the following 
field equations for $\psi(r)$ and $A_0(r) \equiv \phi(r)$, setting in the following the electric charge to unity, $e=1$
\begin{equation}
\psi_{rr} + \left( \frac{f_r}{f} + \frac{D-2}{r}\right) \psi_r + \left( \frac{\phi^2}{f^2} - \frac{m^2}{f} \right) \psi = 0
\end{equation}
\begin{equation}
\phi_{rr} + \left( \frac{D-2}{2q-1} \right) \: \frac{\phi_r}{r} - \frac{2 \psi^2 \phi (\phi_r)^{2 (1-q)}}{q (2q-1) (-2)^{1+q} \beta f} = 0
\end{equation}
subjected to the following boundary conditions: At the event horizon, $r \rightarrow r_H$,
\begin{equation}
\phi(r_H) = 0
\end{equation}
\begin{equation}
\psi(r_H) = \frac{(D-1) r_H}{m^2} \: \psi_r(r_H)
\end{equation}
while at the boundary, $r \rightarrow \infty$, the solutions are required to behave as follows
\begin{equation}
\psi \sim \frac{\psi_-}{r^{\lambda_-}} + \frac{\psi_+}{r^{\lambda_+}}
\end{equation}
\begin{equation}
\phi \sim \mu - \frac{\rho^{\frac{1}{2q-1}}}{r^\beta}
\end{equation}
where the powers $\beta$ and $\lambda_i$ are given by
\begin{equation}
\beta = \frac{D-2}{2q-1} - 1
\end{equation}
\begin{equation}
\lambda_{\pm} = \frac{1}{2} [(D-1) \pm \sqrt{(D-1)^2+4 m^2}]
\end{equation}
respectively, and we may choose either $\psi_-=0$ or $\psi_+=0$. The quantities $\mu$ and $\rho$ are interpreted 
as the chemical potential and the energy density, respectively. In the following we set $\psi_-=0$ and work 
with $\lambda_+$.

\section{Study of the critical temperature: Analytical results}

Since the scalar field plays the role of the order parameter in phase transitions, above the critical temperature 
the scalar field vanishes, and the equation for $\phi$ takes the simple form
\begin{equation}
\phi_{rr} + \left( \frac{D-2}{2q-1} \right) \: \frac{\phi_r}{r} = 0
\end{equation}
The solution that satisfies both the differential equation and the boundary conditions takes the simple form
\begin{equation}
\phi(z) = \xi r_H (1-z^\beta)
\end{equation}
where we have introduced new quantities $z,\xi$ as follows
\begin{equation}
z \equiv \frac{r_H}{r}
\end{equation}
\begin{equation}
\xi \equiv \left( \frac{\rho}{r_H^{D-2}} \right)^{\frac{1}{2q-1}}
\end{equation}

Next, using in the following $z$ as the independent variable instead of $r$, and taking into account the 
boundary condition for the scalar field at the boundary, $z \rightarrow 0$, we set
\begin{equation}
\psi(z) = z^\lambda F(z)
\end{equation}
with some function that satisfies the conditions $F(0)=1$ and $F_z(0)=0$. Then the differential equation for the new function $F(z)$ takes the form
\begin{equation}
F_{zz} + A F_z + B F = - \xi^2 \: \frac{(1-z^\beta)^2}{(1-z^{D-1})^2}  \: F 
\end{equation}
where the coefficients $A(z)$ and $B(z)$ are found to be
\begin{equation}
A(z) = \frac{1}{z} \left( - \frac{2+(D-3) z^{D-1}}{1-z^{D-1}} + 2 \lambda + 4-D \right)
\end{equation}
\begin{equation}
B(z) = - \frac{m^2}{z^2 (1-z^{D-1})} + \frac{\lambda_i (\lambda_i-1)}{z^2} - \frac{\lambda_i}{z^2} \left[D-4+\frac{2+(D-3) z^{D-1}}{1-z^{D-1}}  \right]
\end{equation}

To solve in an analytical way the boundary value problem of the form
\begin{equation}
-(p(z) F_z)' + q(z) F(z) = \xi^2 w(z) F(z)
\end{equation}
in the range $0 \leq z \leq 1$, where $\xi^2$ is the eigenvalue, and $w(z),p(z),q(z)$ are given functions, we try a 
test function $F(z;a)$ with some unknown parameter $a$ that minimizes the expression \cite{siopsis}
\begin{equation}
\xi(a)^2 = \frac{\int_0^1 p(z) [F(z;a)']^2 + \int_0^1 q(z) [F(z;a)]^2}{\int_0^1 w(z) [F(z;a)]^2}
\end{equation}
In our case we use as a test function $F(z)=1-az^2$, and it is easy to verify that the functions $w(z),p(z),q(z)$ are 
given by the following expressions
\begin{equation}
p(z) = (1-z^{D-1}) z^{2 \lambda+2-D}
\end{equation}
\begin{equation}
w(z) = p(z) \: \frac{(1-z^\beta)^2}{(1-z^{D-1})^2}
\end{equation}
\begin{equation}
q(z) = - p(z) B(z)
\end{equation}

Upon minimization of the expression above at $a_*$, we determine the pair of values of $(a_*,\xi_{min})$, and finally 
the critical temperature is given by the corresponding Hawking temperature \cite{NLE1,NLE2}
\begin{equation}
T_H = \frac{(D-1) r_H}{4 \pi}
\end{equation}
or
\begin{equation}
T_c = \frac{D-1}{4 \pi \xi_{min}^{\beta+1}} \: \rho^{\frac{1}{D-2}}
\end{equation}

In the following we shall consider a few special cases, and in particular the case $D=4,q=1$, corresponding to the four-dimensional Maxwell's theory, and the case $D=3,q=3/4$, corresponding to the three-dimensional EpM theory for which the trace of the electromagnetic stress-energy tensor vanishes.

\smallskip

\begin{figure}[ht!]
\centering
\includegraphics[width=\linewidth]{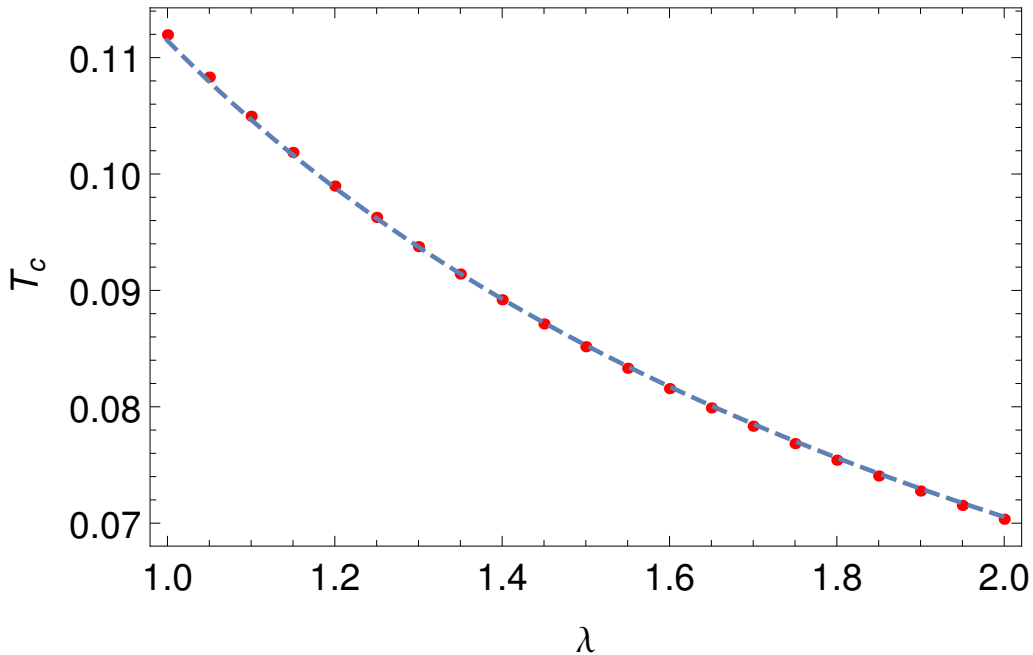}
\caption{Critical temperature (in units of $\rho$) as a function of $\lambda_+$ in the $D=3,q=3/4$ case (EpM theory). 
The dashed curve corresponds to the fitting curve $y=0.111/x^{0.66}$.}
\label{fig:1} 	
\end{figure}

\begin{figure}[ht!]
\centering
\includegraphics[width=\linewidth]{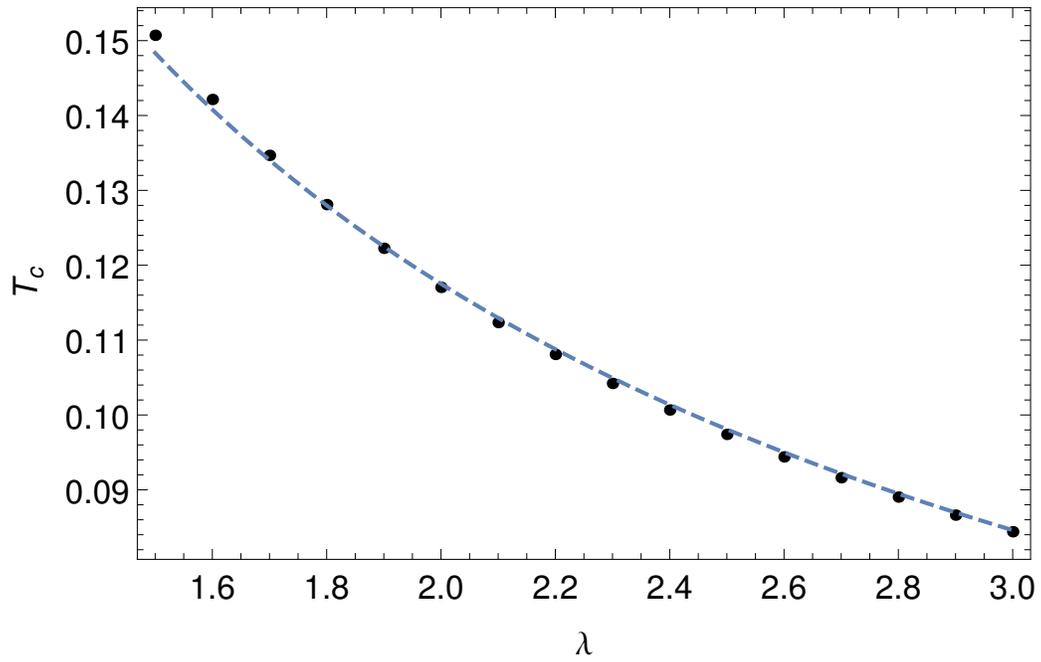}
\caption{Critical temperature (in units of $\rho^{1/2}$) for different values of $\lambda_+$ in the $D=4,q=1$ case (Maxwell's theory). The dashed curve corresponds to the fitting curve $y=0.206/x^{0.81}$.}
\label{fig:2} 	
\end{figure}

\begin{figure}[ht!]
\centering
\includegraphics[width=\linewidth]{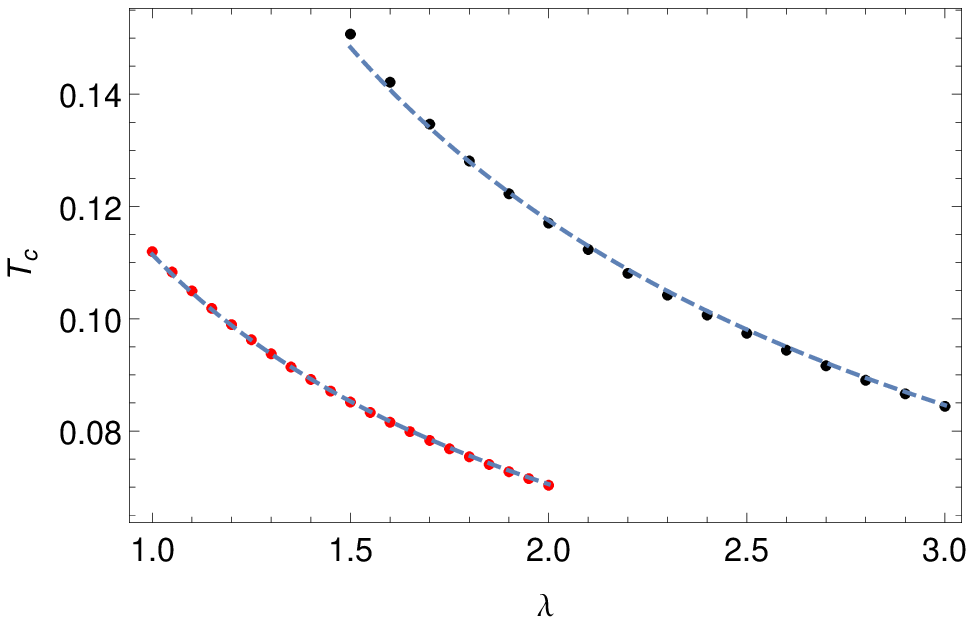}
\caption{Comparison between 1D superconductors in the presence of EpM non-linear electrodynamics (red points), and 2D superconductors in the presence of Maxwell's linear theory (black points).}
\label{fig:3} 	
\end{figure}

Considering a scalar field mass in the range $-1 \leq m^2 \leq 0$, the power $\lambda_+$ takes values in the range $1 \leq \lambda_+  \leq 2$. In Table \ref{table:EpM} we show the critical temperature (for $\rho=1$) for several different values of $\lambda_+$. In the same table we also show $T_c$ for the (1+3)-dimensional Einstein-Maxwell theory. For better visualization, we show our results graphically in Figures~\ref{fig:1} and \ref{fig:2}, and for comparison 
reasons we show them together in the same plot, Fig.~\ref{fig:3}. We also show the fitting curves, 
$y(x)=0.111/x^{0.66}$ for the EpM case, and $y(x)=0.206/x^{0.81}$ for the Maxwell case. Our results show 
that i) the critical temperature decreases with $\lambda_+$, and ii) the critical temperature of one-dimensional superconductors is lower than $T_c$ of two-dimensional superconductors.

\smallskip

Finally, a comparison should be made between the results obtained here, and the results obtained for higher-dimensional
holographic superconductors with a power Maxwell potential studied in previous works. In Tables \ref{table:q1} and \ref{table:q2} we show $T_c$ for various values of the space-time dimensionality, $D$, and the mass of the scalar
field, $m^2$, in EpM theory for $q=3/4$ and $q=5/4$, respectively. The numerical values corresponding to the $D=3$ case 
have been obtained here, whereas the numerical values corresponding to the $D=4,5$ case have been obtained in \cite{NLE1}.
The tabulated numerical values show that the critical temperature increases with $D$ and it decreases with $q$, 
which is in agreement with what is shown in Figures 1 and 2 of \cite{NLE2}.


\begin{table*}[t]
\caption{Critical temperature (setting $\rho=1$) for different values of $\lambda_+$ for the EpM theory (second column) 
and Maxwell's theory (third column).}
{
\begin{tabular}{l | l | l}
		Power $\lambda_+$ & $T_c$ (EpM) & $T_c$ (Maxwell)  \\
		\hline
		\hline
		1.00 & 0.111964 & -\\
		1.05 & 0.108324 & - \\
		1.10 & 0.104964 & - \\
		1.15 & 0.101854 & - \\
		1.20 & 0.0989643 & - \\
		1.25 & 0.0963727 & - \\
		1.30 & 0.0937587 & - \\
		1.35 & 0.0914047 & - \\
		1.40 & 0.0891954 & - \\
		1.45 & 0.0871174 & - \\
		1.50 & 0.085159 & 0.150713 \\
		1.55 & 0.0833099 & - \\
		1.60 & 0.0815606 & 0.142141 \\
		1.65 & 0.0799032 & - \\
		1.70 & 0.0783302 & 0.134672\\
		1.75 & 0.0768351 & - \\
		1.80 & 0.075412  & 0.128097\\
		1.85 & 0.0740555 & - \\
		1.90 & 0.0727609 & 0.122262\\
		1.95 & 0.0715239 & - \\
		2.00 & 0.0703404 & 0.117042 \\
		2.10 & - & 0.112343  \\
		2.20 & - & 0.108087 \\
		2.30 & - & 0.104212 \\
		2.40 & - & 0.100666  \\
		2.50 & - & 0.0974083 \\
		2.60 & - & 0.0944027 \\
		2.70 & - & 0.0916196 \\
		2.80 & - & 0.0890341  \\
		2.90 & - & 0.0866247 \\
		3.00 & - & 0.0843729 \\
\end{tabular}
	\label{table:EpM}
}
\end{table*}


\begin{table*}[t]
\caption{Critical temperature (setting $\rho=1$) for different values of $D,m^2$ in EpM theory with $q=3/4$.}
{
\begin{tabular}{l | l | l | l}
		Mass & $D=3$ & $D=4$ & $D=5$  \\
		\hline
		\hline
		$m^2=-1$ & $T_c=0.1120$ & - & $T_c=0.2545$ \\
		\hline
		$m^2=0$  & $T_c=0.0703$ & $T_c=0.1694$ & $T_c=0.2505$  \\
\end{tabular}
	\label{table:q1}
}
\end{table*}		


\begin{table*}[t]
\caption{Critical temperature (setting $\rho=1$) for different values of $D,m^2$ in EpM theory with $q=5/4$.}
{
\begin{tabular}{l | l | l}
		Mass & $D=3$ & $D=5$   \\
		\hline
		\hline
		$m^2=-1$ & $T_c=0.0357$ & $T_c=0.1065$ \\
		\hline
		$m^2=0$  & $T_c=0.0053$ & $T_c=0.1008$  \\
\end{tabular}
	\label{table:q2}
}
\end{table*}



\section{Conclusions}

To summarize, in the present work we have studied one-dimensional superconductors analyzing 
the dual (1+2)-dimensional gravitational system in the presence of non-linear Electrodynamics. In particular,
we have considered the Einstein-power-Maxwell theory, and we have computed the critical temperature of 
the transition in the probe limit employing the Reyleigh-Ritz variational principle.
After the presentation of the equation for the condensate for generic power $q$ and dimensionality
of space-time $D$, we have focused to the cases $D=3$ and $q=3/4,5/4$, and also to the case
$D=4$ and $q=1$ corresponding to the linear Maxwell's theory in four dimensions for comparison
reasons. The critical temperature as a function of the mass of the scalar field has been obtained, 
and a direct comparison with the (1+3)-dimensional case of the Einstein-Maxwell theory has been made.
We have found that the critical temperature decreases with the mass of the scalar field, and
that $T_c$ of the 4D Maxwell theory lies above the critical temperature corresponding to the
3D EpM theory. Finally, we have made a comparison with two-dimensional holographic superconductors
with a power Maxwell field, considered previously in the literature. We have found that $T_c$
increases with the dimensionality of space-time, and that it decreases with the power $q$, which
is in agreement with results found in older similar works.


\section*{Acknowlegements}

The author wishes to thank the anonymous reviewer for a constructive criticism 
as well as for useful comments and suggestions.
G.~P. thanks the Funda\c c\~ao para a Ci\^encia e Tecnologia (FCT), Portugal, for the 
financial support to the Center for Astrophysics and Gravitation-CENTRA,  Instituto 
Superior T\'ecnico,  Universidade de Lisboa, through the Grant No.~UIDB/00099/2020.


\end{document}